\documentclass[aps,showpacs,twocolumn,twoside]{revtex4}
\usepackage{epsfig}
\begin{document}

\def\ket#1{|#1\rangle}
\def\bra#1{\langle#1|}
\def\scal#1#2{\langle#1|#2\rangle}
\def\matr#1#2#3{\langle#1|#2|#3\rangle}
\def\keti#1{|#1)}
\def\brai#1{(#1|}
\def\scali#1#2{(#1|#2)}
\def\matri#1#2#3{(#1|#2|#3)}
\def\sloup#1#2{\left(\begin{array}{c}#1\\#2\end{array}\right)}
\def\ave#1{\langle #1\rangle}

\title{Evolution of spectral properties along the O(6)-U(5) transition \\ 
       in the interacting boson model. II. Classical trajectories}
\author{
Michal Macek$^1$, Pavel Cejnar$^1$, Jan Jolie$^2$, Stefan Heinze$^2$}

\affiliation{
$^1$Faculty of Mathematics and Physics, Charles University,
V Hole{\v s}ovi{\v c}k{\'a}ch 2, 180\,00 Prague, Czech Republic\\
$^2$Institute of Nuclear Physics, University of Cologne,
Z\"ulpicherstrasse 77, 50937 Cologne, Germany
}

\date{\today}
\begin{abstract}
This article continues our previous study of level dynamics in the
[O(6)$-$U(5)]$\supset$O(5) transition of the interacting boson model 
[nucl-th/0504016] using the semiclassical theory of spectral fluctuations.
We find classical monodromy, related to a singular bundle of orbits 
with infinite period at energy $E=0$, and bifurcations of numerous 
periodic orbits for $E>0$.
The spectrum of allowed ratios of periods associated with $\beta$- and 
$\gamma$-vibrations exhibits an abrupt change around zero energy.
These findings explain anomalous bunching of quantum states in the
$E\approx 0$ region, which is responsible for the redistribution of 
levels between O(6) and U(5) multiplets.
\end{abstract}

\maketitle

\section{Introduction}

In the first part of this work \cite{partI} (hereafter referred
to as Part~I), we have discussed the evolution of level energies and 
wave functions along the [O(6)$-$U(5)]$\supset$O(5) transition in the 
interacting boson model (IBM) \cite{Iach}.
It is known that this transitional class is integrable---due to the O(5) 
underlying symmetry and the associated seniority quantum number $v$---and 
exhibits a second-order ground-state phase transition from deformed 
$\gamma$-soft to spherical equilibrium shapes.

Remind that our family of model Hamiltonians is given by
\begin{equation}
\hat{H}(\eta)=a\left[-\frac{1-\eta}{N^2}(\hat{Q}\cdot\hat{Q})+
\frac{\eta}{N}\,\hat{n}_{\rm d}\right]
\ ,
\label{ham}
\end{equation}
with $\eta\in[0,1]$ denoting a dimensionless control parameter that 
drives the system between the O(6) ($\eta=0$) and U(5) ($\eta=1$) 
dynamical symmetries.
The spectrum of $\hat{H}(\eta)$ at any point of the transitional path
depends on the specific interplay of both terms in Eq.~(\ref{ham}), 
where $\hat{Q}=[s^{\dagger}{\tilde d}+d^{\dagger}{\tilde s}]^{(2)}$ 
represents the quadrupole operator and $\hat{n}_{\rm d}=
(d^{\dagger}\cdot{\tilde d}\,)$ the $d$-boson number operator.
Note that $N$ is the total number of bosons, which in the classical limit 
tends to infinity (both terms in the above Hamiltonian are properly
normalized by the $N^k$ denominators to yield finite contributions in 
this limit), and $a=1$~MeV is an arbitrary scaling factor.
The $N\to\infty$ ground-state shape-phase transitions takes place at 
$\eta_{\rm c}=\frac{4}{5}$.

We have shown that one of the most significant features of spectra in 
the $\eta\in[0,1]$ transitional regime of Hamiltonian 
(\ref{ham}) is the pattern of alternating compressions and dilutions of 
levels with angular momentum $l=0$ around energy $E\approx 0$.
This pattern spreads over a wide interval of the control parameter
between $\eta\approx 0.3$ and 0.8, see Fig.~1 in Part~I.
After deconvoluting spectra with different seniorities, it transforms 
into a sequence of avoided crossings that constitute what we called 
the \lq\lq shock-wave scenario\rq\rq\ \cite{partI}.

The level bunching pattern represents basically a huge oscillation of 
the level density in the $E\approx 0$ region, not dissimilar to shell 
effects in single-particle spectra of some quantum-mechanical 
potentials.
There exists a deep and far-reaching relation between fluctuations of 
the quantum level density and properties of periodic orbits in the 
classical counterpart of the given system \cite{Gut1,Stock}.
While it is known that each periodic orbit brings one oscillatory term 
into the level density, with an amplitude related to the orbit's 
dimensionality and stability \cite{Gut2,Bal,Ber}, the interference of 
several such terms gives rise to spectral {\em beating patterns\/} 
that underlie shell effects in nuclei, quantum dots, or metallic 
clusters \cite{Mot}.
Indeed, as follows from the analysis performed by Balian and Bloch 
\cite{Bal}, the inclusion of just two simplest periodic orbits in 
a spheroidal cavity explains the essentials of the shell structure in 
these systems.

Majority of semiclassical studies on the level-density fluctuations 
was performed for hard-wall systems---two-dimensional billiards or 
three-dimensional cavities \cite{Gut1,Stock}.
In these systems, the calculation is considerably simplified since each 
individual orbit exists with easily predictable properties for all 
energies of the particle bouncing between the walls and contributes
by a well-defined term to the single-particle level density.
Nevertheless, the influence of periodic orbits is equally important 
also in systems with \lq\lq soft\rq\rq\ potentials, where the orbit 
analysis is much more involved.
This is also the case of IBM, where the classical limit for $l=0$ 
describes two-dimensional motions within a bounded (for each finite 
$E$) range of quadrupole deformation parameters, governed 
by a Hamiltonian containing specific kinetic and potential terms
\cite{Hatch,Alha1,Alha2,Alha3}.

The purpose of the present part of our contribution is to show that the 
shell effects and the IBM level bunching phenomenon are indeed of similar 
nature, both originating in some particular features of classical periodic
trajectories.
Nevertheless, our reasoning does not only point to ordinary beating 
patterns, known from hard-wall systems, but makes use of two concepts that 
in the context of nuclear models are somewhat less usual.
The first one relies on {\em bifurcations\/} of periodic orbits \cite{Licht}, 
the second on {\em monodromy\/} in classical and quantum integrable systems
\cite{Cush}.
Both these effects lead to singular contributions to semiclassical trace 
formulas, that provide a simplified description of the level-density 
fluctuations.
Therefore, they can be potentially linked to anomalous effects in quantal 
spectra, such as the level bunching at $E\approx 0$.

The paper is organized as follows: 
In Section \ref{hami} we review the construction of the classical limit 
of the IBM Hamiltonian under study and describe basic features of the 
resulting classical dynamics.
Subsection \ref{berry} briefly recapitulates the Berry-Tabor trace formula 
and the role of singular orbits and bifurcations in the semiclassical 
theory of quantal spectra. 
Numerical analysis of orbits with $l=0$, presented in Sec.~\ref{nume},
shows that in the $E\approx 0$ region our system passes through a robust 
structural change of classical dynamics.
This change is correlated with the occurrence of a singular bundle of $E=0$ 
trajectories and triggers multiple bifurcations of orbits 
in the region $E>0$.
The relation of these findings to the concept of monodromy is discussed 
in Section~\ref{mono}.
Finally, Section \ref{conclu} contains concluding remarks.

\section{Classical Hamiltonian}
\label{hami}

The classical limit of the IBM can be obtained via the well-known procedure, 
elaborated in detail by Hatch and Levit \cite{Hatch}, and by Alhassid and 
Whelan \cite{Alha2,Alha1,Alha3}.
The procedure makes use of Glauber coherent states $\ket{\alpha}\propto
\exp{(\alpha_ss^{\dagger}+\sum_{\mu}\alpha_{\mu}d_{\mu}^{\dagger})\ket{0}}$ 
with complex time-dependent coefficients 
$\alpha\equiv\{\alpha_s,\alpha_{\mu}\}_{\mu=-2,\dots,+2}$ which define 
a set of 12 classical-like variables (both coordinates and momenta).
The equations of motion for $\alpha$ are derived from the time-dependent
variational principle, which results in the Hamilton function
given by the coherent-state average $H_{\rm cl}(\eta;\alpha)=\matr{\alpha}
{\hat{H}(\eta)}{\alpha}$.
This function and analogous counterparts of other operators can be 
obtained by substitutions $s,d_{\mu}\mapsto\alpha_s,\alpha_{\mu}$ and 
$s^{\dagger},d_{\mu}^{\dagger}\mapsto\alpha_s^*,\alpha_{\mu}^*$ in
the respective normal-ordered quantal expressions.

Since Glauber coherent states do not fix the total number of bosons,
an additional constraint must be required, namely $\matr{\alpha}{\hat{N}}
{\alpha}=|\alpha_s|^2+\sum_{\mu}|\alpha_{\mu}|^2=N$.
This (plus an arbitrary choice of the overall phase, $\alpha_s=
\sqrt{N-\sum_{\mu}|\alpha_{\mu}|^2}$) reduces the number of relevant 
degrees of freedom from six to five.
Naturally, the classicality of coherent states becomes more and more 
pronounced as $N$ increases and the fully classical limit is obtained 
in the $N\to\infty$ limit.
To prevent divergence of the corresponding averages, one has to scale 
all operators according to their order [see the $\frac{1}{N^k}$ 
factors in Hamiltonian (\ref{ham})] and to absorb the respective 
factors into the definition of $\alpha$'s.
This leads to the substitution $\alpha_{\mu}\mapsto{\tilde\alpha}_{\mu}
=\frac{\alpha_{\mu}}{\sqrt{N}}$ while, simultaneously, the 
$\frac{1}{N^k}$ factors drop out.

Final expressions for the classical-limit observables are obtained after 
the identification of real coordinates $q_{\mu}$ and momenta $p_{\mu}$ 
via relations $\sqrt{2}{\tilde\alpha}_{\mu}=(-)^{\mu}q_{-\mu}+ip_{\mu}$ 
and $\sqrt{2}{\tilde\alpha}_{\mu}^*=q_{\mu}-(-)^{\mu}ip_{-\mu}$.
The coordinates $q_{\mu}$ are associated with the geometric variables 
describing an instantenous quadrupole deformation of the nucleus and 
its orientation in the laboratory frame.
Due to the fixed boson number average, the motion is constrained by
the condition
\begin{equation}
\sum_{\mu}(p_{\mu}^2+q_{\mu}^2)\leq 2
\label{inter}
\end{equation}
to the interior of a sphere in the 10-dimensional phase space.

\begin{figure}
\epsfig{file=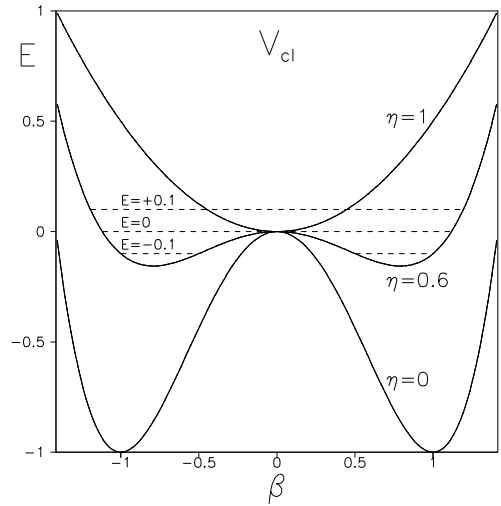,width=6.5cm}
\caption{\protect\small Potential energy term of Hamiltonian (\ref{clas}) for 
three values of parameter $\eta$. The lowermost ($\eta=0$) and uppermost 
($\eta=1$) curves correspond to O(6) and U(5) limits, respectively. The 
middle curve ($\eta=0.6$) represents an intermediate case, for which the 
accessible range of radii (for the three given energies) is shown by the 
dashed lines. \lq\lq Negative radii\rq\rq\ express the $(\beta,\gamma)\to
(\beta,\gamma+180^{\circ})\equiv(-\beta,\gamma)$ transformation and are included 
just to emphasize the rotational symmetry.}
\label{champ}
\end{figure}

The calculation of classical observables is substantially simplified for
zero angular momentum, $l=0$ \cite{Alha3,Str}.
In this case, the intrinsic frame connected with the ellipsoid of
deformation remains at rest and one can fix ${\rm Re}\,q_0\equiv x$, 
and ${\rm Re}\,q_{+2}={\rm Re}\,q_{-2}\equiv y/\sqrt{2}$ (while 
$q_{\pm 1}={\rm Im}\,q_{\pm 2}={\rm Im}\,q_0=0$).
The $l=0$ classical limit of Hamiltonian (\ref{ham}) reads as
\begin{equation}
H_{\rm cl}=
\underbrace{\frac{\eta}{2}\,\pi^2+(1-\eta)\,\beta^2\pi^2}_{T_{\rm cl}}+
\underbrace{\frac{5\eta-4}{2}\,\beta^2+(1-\eta)\,\beta^4}_{V_{\rm cl}}\ ,
\label{clas}
\end{equation}
where $\beta^2=x^2+y^2$ is the squared radius in the $q_0\times\sqrt{2}
q_{\pm 2}$ plane (the polar angle denoted as $\gamma$) and $\pi^2$ the 
squared length of the associated vector of momenta:
\begin{equation}
\pi^2=\pi_x^2+\pi_y^2=
\pi^2_{\beta}+\biggl(\frac{\pi_{\gamma}}{\beta}\biggr)^2
\ .
\end{equation}

Hamiltonian (\ref{clas}) can be thought to describe planar motions of 
a particle with the position-dependent kinetic energy $T_{\rm cl}$ in 
potential $V_{\rm cl}$, which is for $\eta=0$, 0.6 and 1 shown in 
Figure~\ref{champ}.
While for $\eta<\eta_{\rm c}=\frac{4}{5}$, the potential has the 
\lq\lq Mexican hat\rq\rq\ (or \lq\lq champagne bottle\rq\rq) form, 
for $\eta\geq\eta_{\rm c}$ it is just a well with minimum at $\beta=0$. 
To emphasize the rotational symmetry in the $x\times y$ plane, we 
show both positive and \lq\lq negative\rq\rq\ domains of $\beta$ (the 
latter corresponding to the rotation by angle $180^{\circ}$).
As follows from the form of the potential and from Eq.~(\ref{inter}), the 
radius must satisfy
\begin{equation}
\beta \in  [\beta_{\rm min},\beta_{\rm max}]\subset[0,\sqrt{2}]
\label{constr0}
\end{equation}
and the total energy
\begin{eqnarray}
E \in  [E_{\rm min},E_{\rm max}]\subset[-1,+1]
\ ,\qquad\qquad
\label{constr}\\
E_{\rm min}  =  \left\{ \begin{array}{ll}
-\frac{(5\eta-4)^2}{16(1-\eta)} & {\rm for\ }\eta<\frac{4}{5}\\
0 & {\rm for\ }\eta\geq\frac{4}{5}
\end{array}\right.
\quad,\quad
E_{\rm max}=\eta
\ .
\nonumber
\end{eqnarray}
Remind that throughout this paper the energy is always expressed in 
units of the scaling constant $a$, see Eq.~(\ref{ham}), so it is 
formally dimensionless.

Note that polar coordinates $\beta$ and $\gamma$ of $x$ and $y$ can 
be immediately associated with Bohr geometric variables, but in this 
case the deformation parameter would be restricted to $\beta\in
[0,\sqrt{2})$.
To obtain ${\tilde\beta}\in[0,\infty)$, as is usual in nuclear
structure, the coordinate 
plane must be radially stretched \cite{Klein} according to 
$\beta\mapsto{\tilde\beta}=\frac{\beta}{\sqrt{2-\beta^2}}$. 
Expressed in new polar coordinates $({\tilde\beta},\gamma)$ and the 
associated momenta $({\tilde\pi}_{\beta},\pi_{\gamma})$, Hamiltonian 
(\ref{clas}) transforms into the following form:
\begin{eqnarray}
H_{\rm cl}=
\frac{1}{4}\left[\eta+(4-3\eta){\tilde\beta}^2\right]
\left[(1+{\tilde\beta}^2)^2{\tilde\pi}_{\beta}^2+\left(
\frac{\pi_{\gamma}}{{\tilde\beta}}\right)^2\right]
\nonumber\\
+\frac{(5\eta-4){\tilde\beta}^2+\eta{\tilde\beta}^4}
{(1+{\tilde\beta}^2)^2}\ .
\qquad
\label{clas2}
\end{eqnarray}
Here, the upper physical limit of energy, $E_{\rm max}$, is reached when 
the motion becomes infinite.
In the following, we will use the classical limit in the form
of Eq.~(\ref{clas}), i.e., with $\beta$ restricted to the finite
interval (\ref{constr0}).

It is immediately apparent that the Hamiltonian in Eqs.~(\ref{clas}) 
and (\ref{clas2}) is \lq\lq$\gamma$-soft\rq\rq, invariant under rotations 
about the origin, so it conserves the \lq\lq angular momentum\rq\rq
\begin{equation}
\pi_{\gamma}=x\pi_y-y\pi_x\ .
\label{angul}
\end{equation}
Thus, since the number of degrees of freedom $f=2$, the system must be 
integrable.
This is in agreement with the arguments explaining the integrability 
of the [O(6)$-$U(5)]$\supset$O(5) Hamiltonians with arbitrary angular
momenta, as outlined in Part~I 
\cite{partI}, since Eq.~(\ref{angul}) is closely related to the integral 
of motion $\hat{C}_2[{\rm O}(5)]=\frac{1}{5}(\hat{L}\cdot\hat{L})+
2(\hat{T}_3\cdot\hat{T}_3)$.
Indeed, for $l=0$ the classical limit of the O(5) Casimir invariant 
reads as \cite{Hatch}: 
\begin{equation}
C_2[{\rm O}(5)]_{\rm cl}\biggr|_{l=0}=2\pi_{\gamma}^2\ .
\label{senior}
\end{equation}
Note that since \lq\lq angular momentum\rq\rq\ (\ref{senior}) does not
correspond to ordinary O(2) algebra of two-dimensional rotations, its 
quantization yields eigenvalues $v(v+3)$, where for $l=0$ the seniority
takes values $v=0,3,6,\dots$, in contrast to the $m^2$ formula with
$m=0,\pm 1,\pm 2,\dots$ corresponding to O(2).
Nevertheless, we realize that each value of $C_2[{\rm O}(5)]_{\rm cl}$ 
is associated with both signs of $\pi_{\gamma}$, i.e., with two opposite 
orientations of the motion in $\gamma$-direction.
This intrinsic \lq\lq degeneracy\rq\rq\ (which does not affect physical
results in the quantum case) will become important in Sec.~\ref{mono}.

\begin{figure}
\epsfig{file=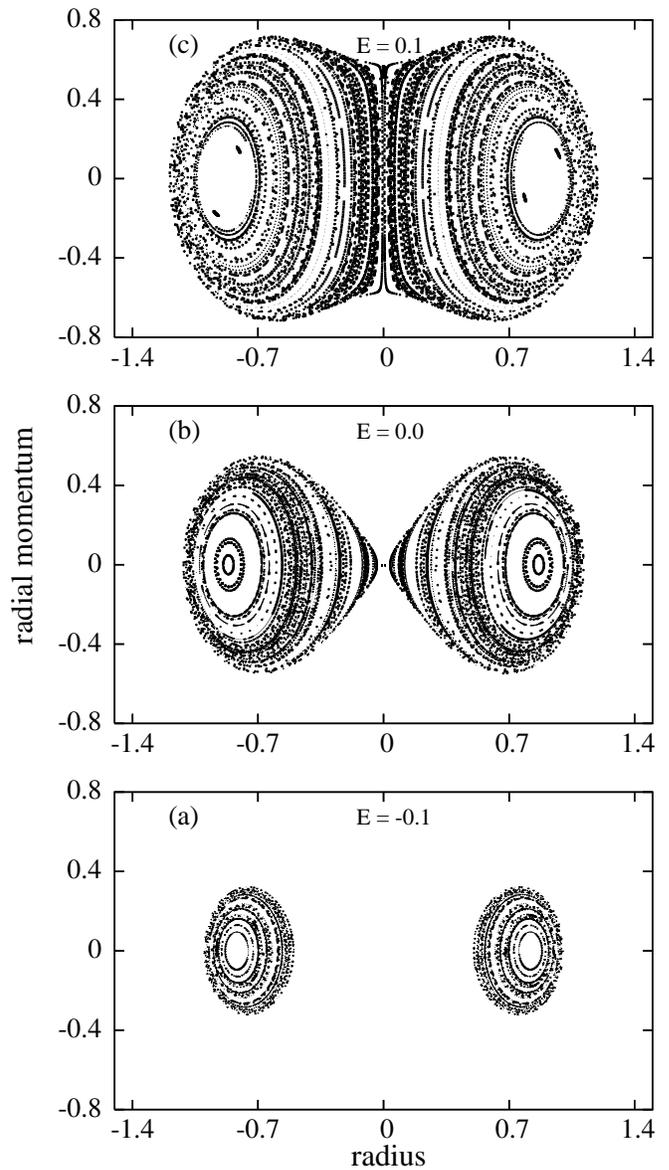,width=\linewidth}
\caption{\protect\small Poincar{\' e} phase-space sections for Hamiltonian
(\ref{clas}) with $\eta=0.6$ at the three given values of energy (panels
a--c). The sections show a finite number of crossings of 10 (a), 30 (b), 
and 50 (c) trajectories 
with the plane $\beta\times\pi_{\beta}$ for arbitrary~$\gamma$.}
\label{rezy}
\end{figure}

The integrability of Hamiltonian (\ref{clas}) is illustrated in 
Figure~\ref{rezy}, where we show Poincar{\' e} phase-space sections for 
$\eta=0.6$ at three different energies, (a) $E=-0.1$, (b) $E=0$, and 
(c) $E=0.1$.
Each of the panels represent passages of 10--50 randomly selected 
trajectories 
with the given energy $E$ through the $\beta\times\pi_{\beta}$ plane in 
the 4-dimensional phase space.
Due to the rotational symmetry, the plane can have an arbitrary 
orientation in the $x\times y$ frame and the pattern of sections must 
be symmetric under the reflection of the $\beta$ axis (we show both 
$\beta>0$ and \lq\lq $\beta<0$\rq\rq\ halves).

All sections in Fig.~\ref{rezy} demonstrate fully regular dynamics, in 
agreement with the integrability of our system.
As can be anticipated from Fig.~\ref{champ}, the $E<0$ motions in panel 
(a) must be confined inside the annular region $\beta\in[\beta_{\rm min},
\beta_{\rm max}]$, while the $E>0$ trajectories in panel (c) already range 
over the full disc $\beta\in[0,\beta_{\rm max}]$ (the values $\beta_{\rm min}$ 
and $\beta_{\rm max}$ depend on energy).
Panel (b) shows just the singular $E=0$ situation, when the central inaccessible
(for $E<0$) disc shrinks into a single point (which can be reached in infinite
time).
We will return to this case in Sec.~\ref{mono}.

Let us stress that the Poincar{\' e} sections in Fig.~\ref{rezy} separate
trajectories with different energies, but mix together those with
various values of the angular momentum $\pi_{\gamma}$.
Indeed, the outermost curves in all panels represent pure $\beta$-vibrations 
with $\pi_{\gamma}=0$, while the central points (not shown) correspond to 
\lq\lq spinning\rq\rq\ only in the $\gamma$-direction with $\pi_{\beta}=0$.
The other trajectories correspond to various mixtures of $\beta$- and 
$\gamma$-vibrations.
An interesting attribute of these intermediate cases is the spread 
$\Delta\beta$ of each individual trajectory
in the $\beta$-direction, which can be determined as the 
difference between radii corresponding to the outermost and innermost points.
This will be used in Sec.~\ref{orbits} to classify quasiperiodic orbits 
in our system.

\section{Periodic orbits}
\label{orbits}

\subsection{Berry-Tabor formula, singular orbits, and bifurcations}
\label{berry}

Semiclassical analyses of quantal spectra are performed in the framework 
of so-called trace formulas which represent the fluctuating part 
$\varrho_{\rm fl}(E)$ of the level density $\varrho_{\rm tot}=\varrho_{\rm sm}
+\varrho_{\rm fl}$ in terms of purely classical quantities associated 
with periodic orbits, while the complementary smooth part $\varrho_{\rm sm}(E)$
is determined just as the classical phase-space volume available at a given 
energy $E$ in units of $\hbar^f$ (where $f$ is the number of system's degrees 
of freedom).
The best known expression, derived by Gutzwiller \cite{Gut2}, was obtained 
under the assumption that individual periodic orbits are isolated, which is 
not satisfied for integrable systems.
In this case, periodic orbits come in continuous families characterized 
by arbitrary shifts of initial angles if the motion is described in the 
action-angle variables \cite{Licht}.
An adequate semiclassical approach to the level density of integrable
systems was developed by Berry and Tabor \cite{Ber}.

In the following, we consider a two-dimensional integrable system, 
$f=2$, which applies in our case of Hamiltonian (\ref{clas}).
In the action-angle representation the Hamiltonian depends only on 
actions, $H=H(I_1,I_2)$, and the angles evolve according to 
${\dot\theta_1}=\omega_1$ and ${\dot\theta_2}=\omega_2$.
All trajectories represent folded rotations on various tori determined
by ${\vec I}=(I_1,I_2)$.
Let us note that in many integrable systems, including ours, the 
action-angle variables can only be introduced locally \cite{Cush} 
(cf. Sec.~\ref{mono}).
Therefore, it is not possible to write down analytic expressions for the 
corresponding canonical transformation from normal coordinates and 
momenta.

Any primitive periodic orbit on a given torus can be characterized by 
a pair of coprime integers $(\mu_1,\mu_2)\equiv{\vec\mu}$ such that the 
ratio of angular frequencies ${\cal R}=\frac{\omega_1}{\omega_2}$ 
coincides with the 
rational number $\frac{\mu_1}{\mu_2}$.
The Berry-Tabor formula for the fluctuating part of the quantal state 
density \cite{Ber} then reads as
\begin{eqnarray}
\varrho_{\rm fl}(E)=\frac{1}{\pi\hbar}\sum_{{\vec\mu}}\sum_{r=1}
^{\infty}\frac{T_{{\vec\mu}}}{\sqrt{\hbar|g''_E|(r\mu_2)^3}}
\qquad\qquad\qquad
\nonumber\\
\times
\cos\left[\frac{1}{\hbar}rS_{{\vec\mu}}(E)-\frac{\pi}{2}r
\nu_{{\vec\mu}}-\frac{\pi}{4}\right]\ ,
\label{BT}
\end{eqnarray}
where the sum runs over all repetitions $r$ of all primitive orbits
${\vec\mu}$ with period $T_{{\vec\mu}}=\frac{2\pi\mu_1}{\omega_1}=
\frac{2\pi\mu_2}{\omega_2}$, Maslov index $\nu_{{\vec\mu}}$ \cite{Gut1}, 
and action 
\begin{equation}
S_{{\vec\mu}}(E)=2\pi{\vec I}\cdot{\vec\mu}=\int_0^{T_{\vec\mu}}
[\pi_{\beta}\,{\dot\beta}+\pi_{\gamma}\,{\dot\gamma}]\,dt
\ .
\label{action}
\end{equation}
The meaning of the function $g''_E$ in Eq.~(\ref{BT}) will be explained 
later.

Expression (\ref{action}), which in the general case integrates the scalar 
product of momentum and velocity over the specific periodic trajectory 
${\vec\mu}$, has a particularly simple form for billiards (or cavities), 
where one can write $S_{{\vec\mu}}=2ET_{{\vec\mu}}=pL_{{\vec\mu}}$ with 
$p=mv$ denoting the ordinary momentum and $L_{{\vec\mu}}$ the length of 
the given orbit.
For \lq\lq soft\rq\rq\ systems, the dependence of $S_{{\vec\mu}}$ on energy
is nonlinear and frequencies of individual cosine terms in 
Eq.~(\ref{BT}) vary with $E$.
Since in the latter case each oscillatory term in the Berry-Tabor 
formula contains also a nontrivial energy dependence of the amplitude,
the semiclassical analysis of spectra in such cases is certainly much less 
intuitive than in the hard-wall systems.

In general, there may exist singular orbits with diverging contributions 
to the Berry-Tabor formula. 
This happens if either the period of the given orbit grows to infinity,
$T_{{\vec\mu}}\to\infty$, or if the denominator of the prefactor in 
Eq.~(\ref{BT}) vanishes, $g''_E\to 0$.
The former case applies to the motions that for some energy become 
infinitely slow at a certain point, which can be associated with an 
unstable equilibrium of the system.
We know from the discussion in Sec.~\ref{hami} that our system contains
such a point, namely the central maximum of the potential in 
Eq.~(\ref{clas}) at $\beta=0$ for $\eta\leq\frac{4}{5}$.
For trajectories with $E=0$, this maximum can only be reached in
asymptotic times because the force vanishes there.
Among the trajectories passing this point there are also various periodic 
orbits, whose contribution to Eq.~(\ref{BT}) must diverge at $E=0$ because 
of the period tending to infinity.
This is essentially the classical mechanism responsible for the bunching 
of quantum levels in the region $E\approx 0$, see Fig.~1 in Part~I
\cite{partI}.
It shows that the bunching pattern is not just a finite-$N$ quantum 
fluctuation, but a robust effect deeply ingrained in the classical limit 
of the system.
Theoretical foundations underlying the existence of the singular class
of trajectories and another approach to understand their influence on 
the quantum spectrum will be discussed in Sec.~\ref{mono}.

The second possible source of infinite contributions to the Berry-Tabor
formula (\ref{BT})
is connected with the cases when $g''_E(I_1)\equiv\frac{\partial^2g_E}
{\partial I_1^2}(I_1)=0$.
The function $g_E(I_1)$ is determined \cite{Boh} from the implicit 
equation $H(I_1,I_2=g_E)=E$, which after differentiation and the use 
of Hamilton equations yields 
\begin{equation}
{\dot\theta_1}+{\dot\theta_2}\frac{\partial g_E}{\partial I_1}=0\ ,
\end{equation} 
so that $g_E'=-\frac{\omega_1}{\omega_2}=-{\cal R}$.
In other words, the function $g_E$ matches possible pairs of actions 
$(I_1,I_2)$, i.e., 
selects the tori ${\vec I}$ relevant at a given energy, and its first 
derivative determines the corresponding frequency ratios.
If $-g_E'$ is rational, for a selected torus, the associated orbit is periodic 
and contributes to Eq.~(\ref{BT}).
The second derivative $g_E''$ measures the change of ${\cal R}$ as one steps 
to the tori in an infinitesimal vicinity of ${\vec I}$. 
If $g_E''\neq 0$, the periodic orbit ${\vec\mu}$ on the torus ${\vec I}$
does not survive the transition to ${\vec I}+\vec{\delta I}$.
If, however, $g_E''=0$, a family of periodic orbits with the same frequency
ratio ${\cal R}$ 
exists in neighboring tori, which results in diverging contribution to the 
Berry-Tabor formula (\ref{BT}).

Note that the Gutzwiller formula \cite{Gut2}, which is valid in nonintegrable 
systems with isolated periodic orbits, is formally similar to Eq.~(\ref{BT}), 
but with the prefactor denominator replaced by $\sqrt{\det[(M_p)^r-1]}$, 
where $M_p$ stands for the so-called monodromy matrix of a given primitive 
periodic orbit $p$ \cite{Gut1,Stock}.
This matrix describes the stability of orbit $p$ in terms of 
linearized deviations from the given phase-space trajectory under 
a perpendicular perturbation of its initial point.
Thus $(M_p)^r-1$ represents the deviation from the perturbed phase-space
position after $r$ repetitions.
If one (or more) of the eigenvalues of this matrix is equal to zero, i.e., 
if $\det[(M_p)^r-1]=0$, there exists at least one direction in the phase
space in which any deviation from the given orbit $r\cdot p$ results in 
another periodic orbit.
The new orbits are detached from the primitive orbit $p$ as its period 
$r$-tupling clones.  
Consequently, $p$ is not isolated and the corresponding term in the 
Gutzwiller formula diverges.
This situation is analogous to the one with $g_E''=0$, as described 
above.

Both the above singular cases correspond to the same 
general phenomenon, called bifurcation \cite{Nicol}.
In Hamilton systems of classical mechanics, bifurcations represent 
branching of periodic orbits at some critical values of energy or
other parameters \cite{Licht}.
While periodic orbits existing below and above the given bifurcation
energy $E_{\rm b}$ are isolated, either in the sense of $g_E''\neq 0$ 
or $\det[(M_p)^r-1]\neq 0$, at $E=E_{\rm b}$ two or more orbits merge 
together in the way described above, giving rise to zero denominators 
of the respective semiclassical level-density formulas.
At the bifurcation energies $E_{\rm b}$, the Berry-Tabor or Gutzwiller 
formulas do not represent correct approximations of the fluctuating level 
density. 
Improved semiclassical methods were developed to treat these situations 
\cite{Sie}.
Intuitively one expects an enhancement of the level-density oscillations 
at the bifurcation points.
In the following subsection we will show that in our system numerous 
bifurcations of periodic orbits take place in the energy range $E>0$.

\subsection{Numerical results}
\label{nume}

We have performed a numerical analysis of classical motions corresponding 
to Hamiltonian (\ref{clas}) in the interval of energies $E\in[-0.1,+0.3]$
using a sample of about 50000 generated orbits.
Individual trajectories were calculated with initial positions and momenta 
chosen randomly within the phase-space region accessible at a given energy 
and classified by the ratio 
\begin{equation}
R=\frac{T_{\gamma}}{T_{\beta}}=
\frac{\ave{\omega_{\beta}}}{\ave{\omega_{\gamma}}}
\label{fract}
\end{equation}
of periods $T_{\gamma}$ and $T_{\beta}$ associated with oscillations in 
both $\gamma$ and $\beta$ directions, respectively.
Since $\omega_{\gamma}={\dot\gamma}$ and the angular velocity 
$\omega_{\beta}$ connected with $\beta$-vibrations are both time dependent, 
one has to use the corresponding average angular frequencies per period,
$\ave{\omega_{\gamma}}=\frac{2\pi}{T_{\gamma}}$ and 
$\ave{\omega_{\beta}}=\frac{2\pi}{T_{\beta}}$.
Their inverse ratio coincides with $R$ and is analogous to the above-discussed 
ratio ${\cal R}=\frac{\omega_1}{\omega_2}$ of frequencies in the action-angle 
variables.
In particular, rational values $R=\frac{\mu_{\beta}}{\mu_{\gamma}}$ 
correspond to periodic orbits with period $T_{\mu_{\beta}/\mu_{\gamma}}=
\mu_{\beta}T_{\beta}=\mu_{\gamma}T_{\gamma}$.

\begin{figure}
\epsfig{file=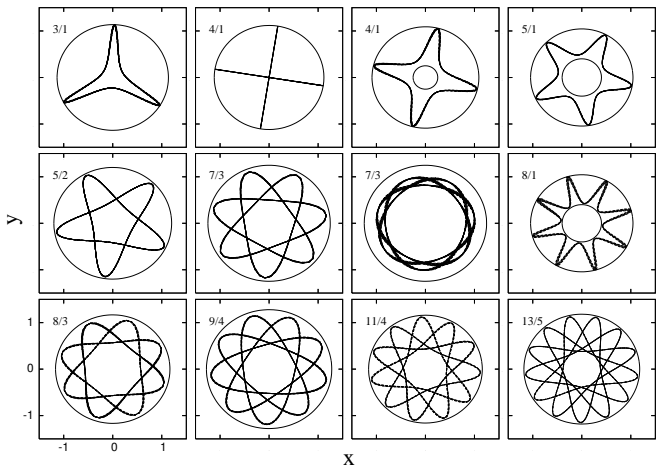,width=\linewidth}
\caption{\protect\small Examples of various periodic orbits at different
energies for Hamiltonian (\ref{clas}) with $\eta=0.6$ and their 
classification by rational fractions $R=\frac{\mu_{\beta}}{\mu_{\gamma}}$.}
\label{traje}
\end{figure}

Examples of periodic orbits with various rational values of the ratio 
(\ref{fract}) are shown in Figure~\ref{traje}.
The rational fraction $R=\frac{\mu_{\beta}}{\mu_{\gamma}}$ classifying the
given orbit has visual meaning as the number of outer return points 
of the $\beta$-vibration over the number of rotations in the
$\gamma$-direction needed to close the orbit.
Thus, for instance, the $5/2$ orbits look like stars with 5 outer 
\lq\lq points\rq\rq\ that close in 2 rotations, while the $5/1$ stars are 
similar, but close only in 1 rotation.
The outer and in some panels also the inner circles in Fig.~\ref{traje} 
demarcate the energetically accessible areas $\beta\in[\beta_{\rm min},
\beta_{\rm max}]$ in the $x\times y$ plane.
As discussed in Sec.~\ref{hami}, this area is a disc for $E>0$, 
an annular ring for $E<0$, and a disc minus the central point for $E=0$.
We see that although the orbits in Fig.~\ref{traje} do not just trivially 
bounce between the outer (and inner) limits, as in the case of 
circular or annular infinite sharp wells, they still resemble to a large 
extent the trajectories in these simple systems \cite{Rob}.

\begin{figure}
\epsfig{file=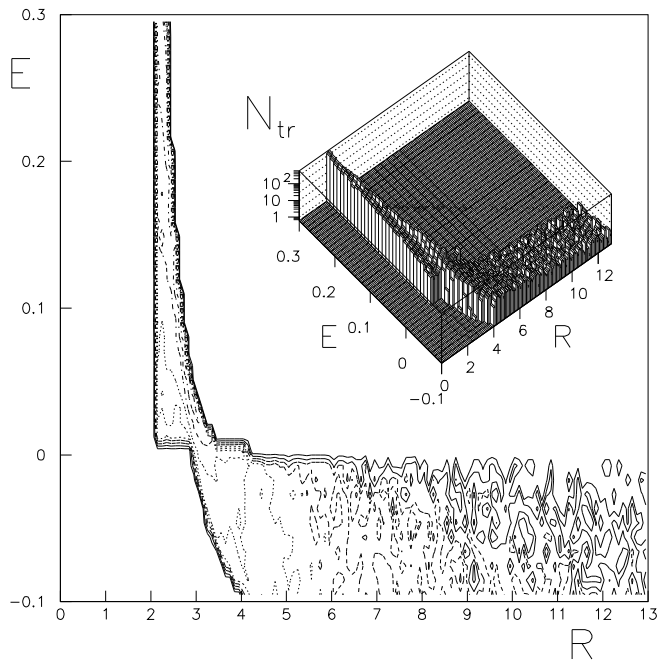,width=\linewidth}
\caption{\protect\small Frequency of occurrence of trajectories with different 
ratios $R$ for Hamiltonian (\ref{clas}), $\eta=0.6$, 
in the reference sample of trajectories with different energies. For each 
value of $E$ (step $\Delta E=0.01$) there was 1200 generated trajectories 
and the histogram (see the inset) shows their distribution in $R$ (the bin 
width $\Delta R=0.01$). The main diagram (contour plot of the logarithmic 
histogram) depicts the band structure of allowed $R$ values, see 
Eq.~(\ref{band}) and below.}
\label{domain}
\end{figure}

Periodic orbits form a dense subset of all allowed motions and we therefore
need a more complete picture.
A histogram showing the occurrence of trajectories with arbitrary (rational
or irrational) values of the frequency ratio $R$ within the whole sample of 
trajectories with $E\in[-0.1,+0.3]$ is presented in Figure~\ref{domain}.
For each value of energy within the given range (the energy step was chosen 
$\Delta E=0.01$), the sample contains $N_{\rm tot}=1200$ trajectories and
Fig.~\ref{domain} depicts their distribution (numbers $N_{\rm tr}$ of
trajectories) into bins of size $\Delta R=0.01$ along the $R$-axis.

The structure shown in Fig.~\ref{domain} discloses rather interesting 
features of classical motions.
For each energy, the orbits occur within a band 
\begin{equation}
R\in[R_{\rm min}(E),R_{\rm max}(E)]
\label{band}
\end{equation}
of allowed frequency ratios.
The lower bound $R_{\rm min}(E)$ gradually decreases with increasing 
energy for $E<0$, but it is constant, $R_{\rm min}=2$, for $E>0$.
Because of the limited precision inherent in our generated sample 
of trajectories (with nonzero values of the bin size $\Delta R$ and 
energy step $\Delta E$) we cannot resolve whether the $R_{\rm min}(E)$ 
dependence is discontinuous or just nonanalytic at $E=0$.
In any case, the curve reaches the minimal value at this point.
On the other hand, for energy approaching the minimum $E_0$ of the
potential in Eq.~(\ref{clas}) (this energy is below the range displayed
in Fig.~\ref{domain}) we must have $R_{\rm min}\to\infty$.
The upper bound of interval (\ref{band}) is also a decreasing function
of energy which passes the value $R_{\rm max}=4$ at $E=0$.
The decrease of $R_{\rm max}(E)$ for $E<0$ (and partly also just above 
$E=0$) is so steep that it cannot be resolved with the present energy 
step, but we assume that it is a smooth curve.
It is obvious that a very narrow energy interval around the point $E=0$ 
carries the most substantial changes in the spectrum of orbits, where 
the trajectories pass between both negative and positive energy regions 
just through a bottleneck of values $R\in[3,4)$.
While for $E<0$ the orbits look similar to those in the O(6) limit, for 
$E>0$ they already resemble the U(5) limit.

The behavior demonstrated in Fig.~\ref{domain} can be qualitatively
understood from the change of the energetically accessible $x\times y$
area around $E\approx 0$.
The form of annular ring, valid for $E<0$, does not support trajectories 
with $R<3$ since these have to traverse through the central region.
Consequently, these trajectories can only exist for $E>-\varepsilon$,
where $\varepsilon\approx 0.03$.
On the other hand, the central reflecting disc is needed for trajectories 
with $R\geq 4$, which therefore appear only for $E<0$.
One can say that the $R\in[4,\infty)$ trajectories, which are 
\lq\lq bouncing\rq\rq\ between inner and outer circles inside the 
annular region for $E<0$, transform to the straight $R\in[2,3]$ 
trajectories at $E\approx 0$ where the central disc gradually
disappears and the accessible domain of deformation parameters becomes 
simply connected.
Note that the rapidity of changes of classical motions around zero energy
is connected with the fact that for $E\to 0_-$ the radius of the central 
disc converges to zero with a rate increasing to infinity ($\beta_{\rm min}
\propto\sqrt{-E}$), as directly follows from the form of the potential 
$V_{\rm cl}$ close to the $\beta=0$ maximum.

\begin{figure}
\vspace{-2cm}
\begin{flushleft}
\epsfig{file=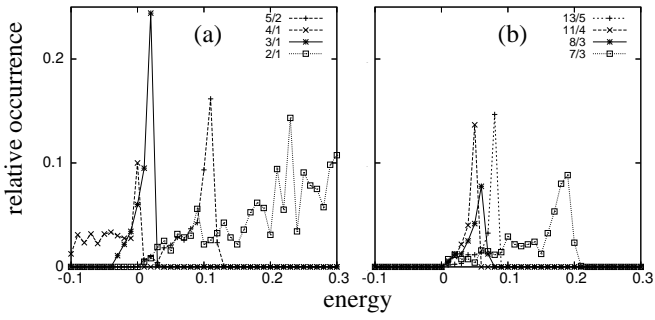,width=\linewidth,angle=-90}
\end{flushleft}
\vspace{-2.5cm}
\caption{\protect\small Relative frequency of occurrence of several types
of quasiperiodic orbits (see Fig.~\ref{traje}) for $\eta=0.6$ as a function 
of energy. The scale on the vertical axis depends on the value
$\Delta\gamma$ in Eq.~(\ref{fractrac}), here equal to $5^{\circ}$.}
\label{occur}
\end{figure}

Figure~\ref{occur} shows the relative frequency of occurrence of several 
types of {\em periodic\/} orbits from Fig.~\ref{traje} in our generated 
sample as a function of energy.
The curves in Fig.~\ref{occur} can be basically understood as energy cuts 
of the function $N_{\rm tr}$ in Fig.~{\ref{domain}} at the respective 
rational values $R=\frac{\mu_{\beta}}{\mu_{\gamma}}$ of the fraction 
(\ref{fract}), but with a variable precision $\Delta R$.
More specifically, Fig.~\ref{occur} presents the relative fraction of all 
generated (at each energy) trajectories satisfying the condition that the 
$\mu_{\beta}$th outer reflection after $\mu_{\gamma}$ revelations is 
shifted from the 1st outer reflection by an angle not exceeding (in
absolute value) the selected precision $\Delta\gamma=5^{\circ}$.
This leads to the condition
\begin{equation}
\frac{\left|R-\frac{\mu_{\beta}}{\mu_{\gamma}}\right|}{R}\leq
\frac{\Delta\gamma}{2\pi\mu_{\gamma}}
\ . 
\label{fractrac}
\end{equation}
Let us stress that the use of a smaller value of $\Delta\gamma$ decreases 
the yield of trajectories---implying a prolongation of the computation 
time---but does not change 
(as we checked for $\Delta\gamma=1^{\circ}$) the shape of dependences in
Fig.~\ref{occur}.

The most common type of behavior shown in Fig.~\ref{occur} indicates that
for many orbits the relative frequency of occurrence sharply culminates at 
a certain energy, just before this orbit totally disappears from the 
system.
The sharpest peak of this kind is observed at $E=0.02$ for the 3/1 orbits, 
but there 
are also other well pronounced peaks, like the 5/2,  11/4, or 13/5 ones,
and many others.
All these maxima appear at positive energies and one can trace their origin 
to the ridge of values $N_{\rm tr}$ visible for $E>0$ at the upper bound 
$R_{\rm max}(E)$ in Fig.~{\ref{domain}} (see the inset).
The peak at $E=0$ (the 4/1 \lq\lq crosses\rq\rq) and also the one at 
$E=0.02$ (the 3/1 \lq\lq Mercedes-Benz stars\rq\rq, see Fig.~\ref{traje}) 
are located just
on the upper edge of the major $E\approx 0$ level bunching pattern in 
Fig.~1 of Part~I \cite{partI}.

Special attention should be paid to the 4/1 orbits that in our system
take two different forms:
For $E<0$ they exist as stars, shown in the third uppermost panel of
Fig.~\ref{traje}, but at $E\approx 0$ they can also look like crosses, 
see the second panel.
(In fact, the latter case exemplifies the above-discussed critical
$E=0$ periodic trajectories with infinite period, as will be further
elaborated in Sec.~\ref{mono}.)
The contributions of these forms to the dependence in Fig.~\ref{occur}(a) 
can be decomposed into a constant step-like function equal to zero for 
$E>0$ (\lq\lq stars\rq\rq) and a sharp peak at $E=0$ (\lq\lq crosses\rq\rq).

We also see in Fig.~\ref{occur}(a) that the 2/1 orbits, which pass via the 
central maximum of the potential and correspond to the $E>0$ edge $R_{\rm min}$ 
in Fig.~\ref{domain}, exhibit a different type of energy dependence than the 
others.
The frequency of occurrence of these orbits is zero at $E\leq 0$ and gradually 
increases (if neglecting fluctuations) with energy $E>0$.

\begin{figure}
\begin{flushleft}
\epsfig{file=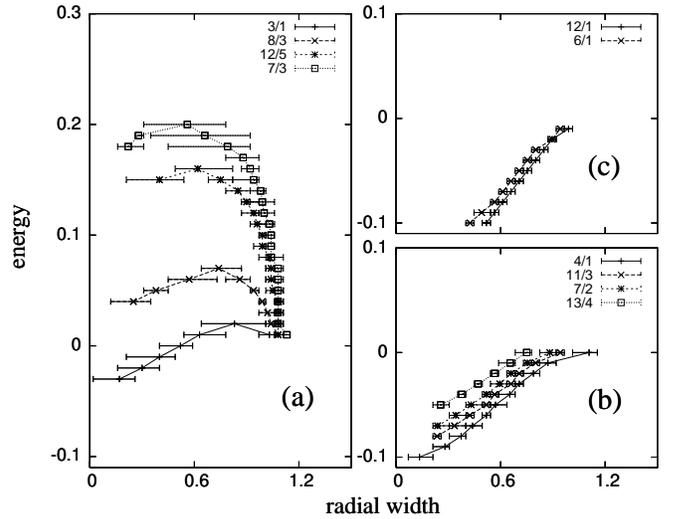,width=6.8cm,angle=-90}
\end{flushleft}
\caption{\protect\small The radial width $\Delta\beta$ of individual 
periodic orbits $\frac{\mu_{\beta}}{\mu_{\gamma}}$ as a function of
energy ($\eta=0.6$).
Horizontal bars at each energy demarcate intervals of $\Delta\beta$ where 
trajectories are detected within our sample, while points connected by 
curves represent statistical averages within each interval.
The bifurcations of orbits in panel (a) explain the respective peaks in 
Fig.~\ref{occur}.}
\label{widle}
\end{figure}

It is not difficult to show that the peaks in Fig.~\ref{occur} corresponding
to the $R\in(2,3]$ orbits are connected with bifurcations.
To this end, we first characterize individual orbits by the radial 
width, defined as the difference $\Delta\beta$ between the outer and inner 
radii (see the end of Sec.~\ref{hami}).
The values of $\Delta\beta$ associated with various orbit types are shown 
in Figure~\ref{widle}, where the horizontal bars demarcate intervals of the
$\Delta\beta$ values which are populated (for energy given on the vertical 
axis) by some trajectories in our sample.
The curves (used just to lead the eye) connect points that represent
arithmetic averages of $\Delta\beta$ in neighboring intervals.

The three panels of Fig.~\ref{widle} collect three types of qualitatively 
different behaviors:
(a) For $R\in(2,3]$, the populated domain of $\Delta\beta$ consists of two 
separate branches (see, e.g., the two 7/3 orbits in Fig.~\ref{traje})
that merge at a certain energy $E_{\rm b}(R)>0$, which 
can be determined from the condition $R_{\rm max}(E_{\rm b})=R$.
Similarly, the lower endpoint energy (the termination of the longer branch 
of the respective curve) follows from the $R_{\rm min}(E)$ bound. 
(b) For $R\in(3,4]$, the domains consist of only one band that shifts 
to larger $\Delta\beta$ values as the energy increases and terminates slightly
above $E=0$. The upper endpoint energies can again be determined from 
$R_{\rm max}(E)$, but as discussed above, this dependence is so steep in 
the given range of $R$ that all endpoint energies in Fig.~\ref{widle}(b) 
fall into the narrow interval $E\in[0,0.01]$. Lower endpoint energies again
follow from $R_{\rm min}(E)$. A special case of this kind is the 4/1 orbit 
with the two above-discussed incarnations (see Fig.~\ref{traje}): the 
respective $\Delta\beta$ value at the endpoint $E=0$ is apparently deviated 
from the direction followed for $E<0$.
(c) For $R>4$, the $\Delta\beta$ bands look similar as in case (b), but
terminate at energies just below $E=0$, supposedly following the steep 
$E<0$ branch of the curve $R_{\rm max}(E)$.

It becomes apparent that the $R\in(2,3]$ peaks in Fig.~\ref{occur} arise 
due to the merge of two different branches of $\Delta\beta$ values, as 
shown in Fig.~\ref{widle}(a).
An increased frequency of occurrence of the orbit just before the endpoint
is related to the flatness of the respective curve close to its maximum 
(a larger number of trajectories is concentrated in a smaller energy 
interval).
At the endpoint $(\Delta\beta_{\rm b},E_{\rm b})$ of each of the curves 
in Fig.~\ref{widle}(a) the respective type of periodic orbit bifurcates, 
having infinitely close neighbors with different radial widths, and thus 
yields $g''_E=0$, as discussed in Sec.~\ref{berry}.
The Berry-Tabor formula (\ref{BT}) cannot be applied at these points
\cite{Sie}.
In our case, the bifurcations seem to be of the pitchfork type, when two 
stable orbits join and produce an unstable one \cite{Nicol}. 
Unfortunately, the unstable orbits are not accessible to numerical 
studies, so they are not seen in Fig.~\ref{widle}.
This problem may be further investigated analytically.

It follows from the above discussion that the bifurcations are connected
only with the region of positive energies (there is a ridge of $N_{\rm tr}$ 
values, apparent in the inset of Fig.~\ref{domain}, which is located solely 
at the $R_{\rm max}(E)$ edge with $E>0$).
This implies that divergences of the Berry-Tabor formula associated with 
bifurcations are not directly relevant in the explanation of the main 
level-bunching pattern in Fig.~1 of Part~I \cite{partI} (except perhaps
the 3/1 case with $E_{\rm b}=0.02$).
Bifurcation energies for low-period orbits are not even correlated with 
the secondary, less pronounced bunchings of levels, observed in the region 
$E>0$ \cite{partI}.
Therefore, it seems that the presence of various orbits in the same energy 
range and an interplay of their bifurcations result in interferences that 
wash out contributions of individual orbits.

On the other hand, highly organized behavior of levels at $E\approx 0$ 
perfectly coincides with the predicted existence of a singular torus of
orbits with infinite period at zero energy, and also with the observed 
abrupt redistribution of the spectrum of orbits in a narrow vicinity of 
this energy.

\section{Classical and quantum monodromy}
\label{mono}

The anomalous $E=0$ bundle of orbits with infinite period, discussed in
Sec.~\ref{berry}, is related to a more general phenomenon, called 
monodromy. 
Classical monodromy in integrable Hamilton systems can be briefly introduced 
as the 
impossibility to define global action-angle variables due to the existence 
of a singular, so called \lq\lq pinched\rq\rq\ torus \cite{Cush}.
The name $M\!o\nu o\delta\!\rho o\mu{\acute\iota}\!\alpha$ (\lq\lq once 
around\rq\rq) originates from a property similar to that of the M{\" o}bius 
strip: if one follows a closed loop in the space of regular tori around the 
singular torus
and---loosely speaking---redefines the coordinate system on the consecutive 
tori continuously on the way along the loop, one returns back to the starting 
torus with a coordinate system that differs from the initial one.

Classical monodromy affects the quantum counterpart of the system via the 
Einstein-Brillouin-Kramers (EBK) quantization rules \cite{Gut1,Stock}.
It turns out that quantum monodromy can be seen as a point defect in the 
lattice of quantum numbers corresponding to a complete set of commuting
operators. 
This defect results in a transformation of the elementary quantum cell 
when a closed loop is completed around the singular point, in analogy
with the above feature of phase-space tori.
An overview of the mathematical background and various examples of
monodromy can be found in Ref.~\cite{Efs}.

Soon after its discovery in 1980 \cite{Dui} it became clear that monodromy
substantially affects global features of numerous integrable systems, which 
might previously be considered as too trivial for detailed analyses.
The simplest system that exhibits monodromy is the spherical 
pendulum---particle moving on a sphere in a gravitational field.
It can be shown \cite{Cush,Efs} that the phase-space torus passing 
the unstable equilibrium position at the north pole, with the particle 
energy exactly equal to the critical value $E_{\rm m}$ needed to reach that 
point, is pinched, i.e., one of its basic circles is contracted to a single
point (with appropriate initial conditions the particle is at rest).
As a consequence, the lattice of quantum states, characterized by quantum
numbers enumerating energy $E$ and the projection $L_z$ of angular momentum,
has a point defect at $(E,L_z)=(E_{\rm m},0)$.
It was found that closely related to this simple observation is the 
realization of monodromy in vibrational and rotational spectra of some 
molecules \cite{Efs,Cush2}.

Other examples of monodromy can be found in the following systems:
particle in quartic, sextic, and decatic potentials \cite{Child}, 
hydrogen atom in orthogonal electric and magnetic fields 
\cite{Cush3}, systems of two or three coupled angular momenta \cite{Sad},
particle bouncing between walls in a prolate elliptic cavity \cite{Waa}
or moving in a two-center attractive potential \cite{Waa2}.
Like in the spherical pendulum, monodromy in several of the latter systems  
is connected with the trajectories passing with the critical energy via
the point of an unstable equilibrium \cite{Efs,Child,Waa2}.
We already know that a similar point, namely the top of the central 
maximum at $\beta=0$, can be found in our classical Hamiltonian 
(\ref{clas}) for $\eta\leq\eta_{\rm c}\equiv\frac{4}{5}$, with 
$E_{\rm m}=0$ being the minimal energy needed to pass through this point.
In fact, our Hamiltonian for $\eta\leq\eta_{\rm c}$ is identical with the 
\lq\lq champagne-bottle\rq\rq\ Hamiltonian of Ref.~\cite{Child}, except 
the position-dependent kinetic term in Eq.~(\ref{clas}), which however 
does not affect the presence of monodromy with the central point
$(E,\pi_{\gamma})=(0,0)$.

\begin{figure}
\epsfig{file=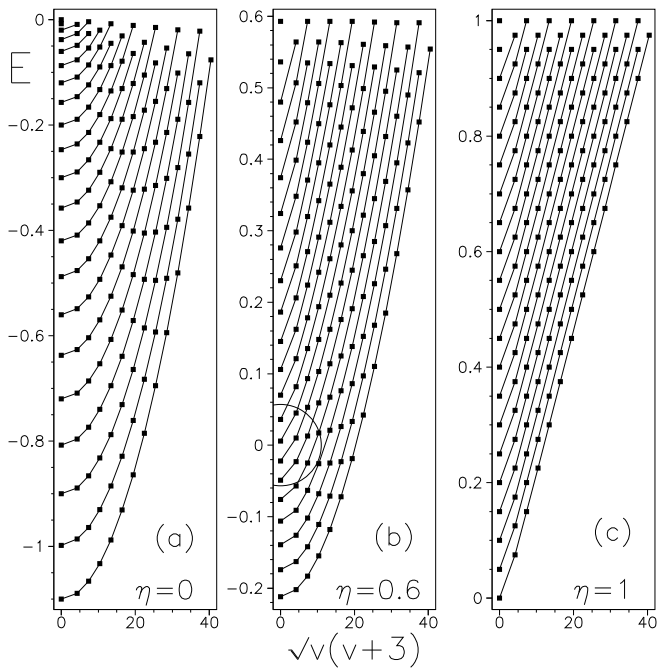,width=\linewidth}
\caption{\protect\small The lattice of $l=0$ eigenstates of Hamiltonian 
(\ref{ham}) with three given values of $\eta$ (panels a--c) and $N=40$ 
in the plane $E\times\sqrt{v(v+3)}$ (where the seniority $v=0,3,6,\dots$). 
Lines connect states with the same radial quantum numbers $n_{\beta}$. 
The singular torus $(E,v)=(0,0)$ is located in the center of the semicircle 
in panel (b).}
\label{lattice}
\end{figure}

Figure~\ref{lattice} shows the lattice of $l=0$ eigenstates of quantum 
Hamiltonian (\ref{ham}) with the number of bosons $N=40$ in the plane where 
the vertical axis represents energy $E$ and the horizontal axis the 
momentum
\begin{equation} 
\pi_{\gamma}\equiv\sqrt{v(v+3)}=3\sqrt{{\tilde v}({\tilde v}+1)}\ .
\end{equation}
Since for zero angular momentum the seniority takes values equal to
multiples of 3, we defined above also the \lq\lq reduced\rq\rq\ 
seniority quantum number ${\tilde v}\equiv \frac{v}{3}=0,1,2,\dots$
The three panels in Fig.~\ref{lattice} correspond to various values 
of the control parameter: (a) $\eta=0$, the O(6) case, (b) $\eta=0.6$, 
a transitional case, and (c) $\eta=1$, the U(5) case.
Individual states (marked by dots) can be directly related to level 
energies at the respective values of $\eta$ in Fig.~1 of Part~I 
\cite{partI}, which collects all states with different seniorities for 
the same boson number as here.

Sorting of states according to seniority in Fig.~\ref{lattice} helps
to identify the values of $v$ that are involved in level bunchings at 
different points $\eta$.
For instance, one immediately sees that the clustering of levels across 
the whole spectrum in the U(5) limit (panel c) is due to the multiple 
degeneracy of states with even or odd values of ${\tilde v}$, that 
correspond to the same value of the U(5) quantum number $n_d$
(even or odd, respectively).
For the highest states, this degeneracy remains approximately 
valid across the whole interval $\eta\in[0,1]$, see panels (a)--(c).
The seniority deconvolution of the spectrum for $\eta<\frac{4}{5}$ is 
exemplified by the $\eta=0.6$ case in panel~(b).
We observe here that levels with all values of ${\tilde v}$ become 
nearly degenerate in the region around zero energy, which is a clear 
signature of the $E\approx 0$ bunching pattern \cite{partI}.

The lattices in Fig.~\ref{lattice} represent quantum energy-momentum maps 
\cite{Efs} of the classical phase space, with each dot being an image of 
a classical torus of trajectories that survived the semiclassical EBK 
quantization \cite{Boh}.
This is given by $I_i=2\pi\hbar(n_i+\frac{\nu_i}{4})$, where $I_i$ (with 
$i=1,2$) are quantized actions and $\nu_i$ the respective Maslov indices 
\cite{Stock}.
The EBK tori should be determined by two quantum numbers $n_1$ and $n_2$,
whose integer values increase by one.
Good candidates for these numbers are the reduced seniority ${\tilde v}$ 
(connecting vertical columns of points in Fig.~\ref{lattice}) and the 
radial quantum number $n_{\beta}=0,1,2,\dots$, that enumerates states 
with a fixed ${\tilde v}$ according to energy (in Fig.~\ref{lattice}, 
the constant-$n_{\beta}$ states are connected by lines).
For $\eta=0$, the radial quantum number is related to $\sigma$, which 
corresponds to the O(6) Casimir invariant \cite{Iach}, and the pair
$(n_{\beta},{\tilde v})$ represents the appropriate choice of the EBK 
quantum numbers.
In the U(5)-like case, as shown below, yet an alternative pair of quantum 
numbers needs to be defined.

It follows from Eq.~(\ref{clas}) that the $l=0$ classical limit of the  
$\eta=1$ Hamiltonian (\ref{ham}) is identical with an isotropic 
two-dimensional harmonic oscillator.
Indeed, for the subset of states with the U(5) quantum number $n_d$ equal 
to multiples of 3 (in this case $n_d=2n_{\beta}+v$) the U(5) lattice 
coincides with the 2D-oscillator lattice of states (the energy in the 
latter case being enumerated by the oscillator quantum number $n_{\rm o}
=2n_r+m$, where $n_r$ and $m$ stand for ordinary radial and 
angular-momentum quantum numbers, respectively).
In the entire U(5) lattice, however, majority of states is located in 
\lq\lq interstitial\rq\rq\ positions with $n_d\neq 3k$; this is because 
the underlying \lq\lq angular-momentum\rq\rq\ algebra differs from the 
ordinary O(2).
Apart from $n_{\beta}$ and ${\tilde v}$, all U(5) states can be labeled 
by a pair of oscillator-like quantum numbers $n_1=n_{\beta}+{\tilde v}$ 
and $n_2=n_{\beta}+2{\tilde v}$.
States with constant values $n_1=0,1,2,\dots$ form upwards inclined 
rows of dots in Fig.~\ref{lattice}(c), while the $n_2=0,1,2,\dots$ 
quantum number connects states in the downwards inclined rows.
The $n_1$ chains are clearly apparent also in both remaining panels (a) 
and (b) of Fig.~\ref{lattice}.

For the purpose of the semiclassical analysis, the lattices in 
Fig.~\ref{lattice} must be extended to cover both positive and 
negative $\pi_{\gamma}$.
Remind from Sec.~\ref{hami} that although the physical quantum states 
can be represented by non-negative values of $\pi_{\gamma}$, the 
intrinsic degeneracy of classical motions in both $\gamma$-directions 
results in mirror imaging of all states with $v>0$ into the 
$\pi_{\gamma}<0$ half-plane (to guarantee a smooth continuation of 
quantum numbers, we assign values ${\tilde v}=-1,-2,\dots$ to these
\lq\lq twin\rq\rq\ states).
In absence of monodromy, one must be able to engage all states in the
{\em extended\/} lattice into a \lq\lq crystal\rq\rq\ grid of continuous 
and {\em smooth\/} lines, corresponding to constant values of two 
compatible global quantum numbers, with \lq\lq elementary cells\rq\rq\ 
of the grid being topologically equivalent to squares.
From Fig.~\ref{lattice}(a) we see that a smooth grid, symmetric 
under the $\pi_{\gamma}\leftrightarrow-\pi_{\gamma}$ reflection, can 
be constructed in the O(6) case, using the pair of generating quantum 
numbers $(n_{\beta},{\tilde v})$.
In the U(5) case (panel c), this choice of quantum numbers produces
a grid of lines that are broken at $\pi_{\gamma}=0$, but a smooth 
global grid (a diagonal \lq\lq chessboard\rq\rq) is generated by the pair 
$(n_1,n_2)$. 
The latter structure can be extended to the whole interval 
$\eta\in[\frac{4}{5},1]$ where the U(5)-like spectrum exists.

In contrast, quantum monodromy implies the absence of a smooth 
{\em global\/} grid.
This is the case of Fig.~\ref{lattice}(b), where a smooth grid for 
$E<0$ would be generated by the pair of quantum numbers 
$(n_{\beta}, {\tilde v})$, but for $E>0$ by the pair 
$(n_1,n_2)$.
Any attempt to define two global quantum numbers that behave 
smoothly in the entire lattice for $\eta\in(0,\frac{4}{5})$ fails at the 
point $(E,v)=(0,0)$, which represents the singular torus of trajectories
and, simultaneously, a \lq\lq defect\rq\rq\ in the quantum lattice of 
states \cite{Efs,Child,Sad}.
It is clear that in the transition to the O(6) limit the defect is 
gradually pushed up to the upper edge of the lattice.
For the whole interval $\eta\in(0,\frac{4}{5})$ the singular point indicates 
the place where the energy-momentum map passes between the O(6) and U(5) 
types of elementary cells---tetragons with ordered $(n_{\beta},{\tilde v})$, 
$(n_{\beta},{\tilde v}+1)$, $(n_{\beta}+1,{\tilde v}+1)$, $(n_{\beta}+1,
{\tilde v})$ vertices, and analogous tetragons in $(n_1,n_2)$, respectively.
Note that elementary cells of either type cannot be uniquely defined along 
a closed loop around the singular point since after one turn the cell 
gets distorted. 
This can be illustrated by a graphical construction in Fig.~\ref{lattice}(b) 
and its mirror image, but the rigorous proof would require an infinite 
density of the lattice in the $N\to\infty$ limit.
The last observation represents a common quantum signature of monodromy
\cite{Efs,Cush2,Child,Cush3,Sad,Waa,Waa2}. 

It should be stressed that monodromy in the present case is not a property of 
just a single Hamiltonian, but characterizes the whole $\eta\in(0,\eta_{\rm c})$ 
family (\ref{ham}) of transitional systems.
Since---as shown in Part~I \cite{partI}---the most substantial changes in 
quantum spectra of these systems take place in the $E\approx 0$ region,
monodromy seems to play the key role in the process of redistribution of 
individual levels between the O(6) and U(5) multiplets.
Related examples exist also in other parametric families of Hamiltonians, 
for instance, in transitions between 
uncoupled and coupled regimes of two quantum rotators \cite{Sad} and between 
Zeeman and Stark limits of the hydrogen atom in crossed electric and magnetic 
fields \cite{Cush3}.
Also in these examples, the crossover between the limiting spectral structures 
takes place at the point (or in the interval) of control parameters and energy 
where monodromy exists.
These findings deserve further investigation.

\section{Conclusions}
\label{conclu}

In the present part of our work, devoted to the [O(6)$-$U(5)]$\supset$O(5)
transition of the interacting boson model, we have studied the classical limit 
of Hamiltonian (\ref{ham}) with zero angular momentum.
Results of the analysis of level dynamics, presented in Part~I \cite{partI}, were 
qualitatively discussed with the aid of the Berry-Tabor semiclassical trace 
formula, which describes fluctuations of quantum spectra in integrable systems 
in terms of families of periodic orbits existing at various energies.
The transitional regime was exemplified by the choice of a single value of
the control parameter, $\eta=0.6$.

Both possible sources of diverging contributions to the trace formula (\ref{BT}), 
namely, the existence of bifurcating ($g''_E=0$) and singular ($T_{{\vec\mu}}=
\infty$) orbits, were identified in our system.
While bifurcations of periodic orbits with ratios between $\gamma$- and 
$\beta$-vibration periods $R\in(2,3]$ were shown to exist in the region $E>0$, 
singular orbits with $\pi_{\gamma}=0$ appear at $E=0$.
The latter finding led to the identification of classical monodromy, which 
on the quantum level exhibits itself as a defect in the lattice of quantum 
states located at zero values of energy and seniority.
This results in the bunching of levels in the $E\approx 0$ region \cite{partI}
and underlies the process of redistribution of states between O(6) and
U(5) spectral structures, i.e., between the $(n_{\beta},{\tilde v})$ and 
$(n_1,n_2)$ types of elementary cells, and the respective multiplets of levels.

Also the numerical analysis of periodic and nonperiodic classical vibrations 
disclosed that the most substantial changes in the spectrum of allowed 
ratios $R$ take place in a very narrow energy interval around $E\approx 0$.
This interval represents a kind of demarcation line between the O(6) and 
U(5) types of classical motions.
At $E\approx-0.03$, the inaccessible central disc in the plane of deformation 
parameters becomes sufficiently small to allow for vibrations with $R\in(2,3]$,
and at $E=0$, when the disc vanishes, the oscillator-like orbits with $R=2$ 
arise.
With the energy further growing to positive values, individual vibrations with 
$R>2$ eventually disappear in bifurcations (\lq\lq annihilations\rq\rq\ of two 
separate $\Delta\beta$ branches of a given orbit). 

We believe that results of this analysis will have a concrete impact on the 
interpretation of data on collective vibrations in $\gamma$-soft nuclei.
In a more general perspective, the [O(6)$-$U(5)]$\supset$O(5) transition of 
the interacting boson model represents a valuable theoretical laboratory for 
studying structural changes between incompatible dynamical symmetries in 
{\em integrable\/} quantum systems.

\acknowledgments
The authors acknowledge important discussions with P.~Leboeuf and
Z.~Pluha{\v r}. This work was supported by the DFG grant no. U36 
TSE 17.2.04.

\end{document}